\documentclass[%
reprint,
prstab,
]{revtex4-1}
\usepackage{amsmath}
\usepackage{amssymb}
\usepackage{graphicx}
\usepackage{dcolumn}
\usepackage{bm}

\begin{document}

\newcommand{\I}{\mathrm{i}}
\newcommand{\E}{\mathrm{e}}
\newcommand{\D}{\,\mathrm{d}}
\newcommand{\s}{\mathrm{sign}}
\newcommand{\sinc}{\mathrm{sinc}}
\renewcommand{\vec}[1]{\bm{\mathrm{#1}}}

\title{Longitudinally strong focusing lattice for Compton--ring gamma sources}
\author{Eugene Bulyak}
\email{bulyak@kipt.kharkov.ua;Eugene.Bulyak@gmail.com}
\altaffiliation[\\Also at: ]{Karazin National University, 4 Svobody sq., Kharkiv, Ukraine}
\affiliation{NSC KIPT, 1 Academichna str., Kharkiv, Ukraine}

\author{Louis Rinolfi}
\affiliation{CERN, Geneva, Switzerland}

\author{Junji Urakawa}
\affiliation{KEK, Ibaraki, Japan}
\begin{abstract}
Electron storage rings of GeV energy with laser pulse stacking
cavities are promising intense sources of polarized hard photons
which, via pair production, can be used to generate polarized
positron beams. Dynamics of electron bunches circulating in a
storage ring and interacting with high-power laser pulses is studied
both analytically and by simulation. Common features and differences
in the behavior of bunches interacting with an extremely high power
laser pulse and with a moderate pulse are discussed. Also
considerations on particular lattice designs for Compton gamma rings
are presented.
\end{abstract}


\maketitle
\section{Introduction. Bottle-necks.}
Compton rings must be able to keep circulating intense electron
bunches with a large spread of energy of individual particles. At
the same time effective generation of the high energy photon beams
demands the bunch length as short as possible.

One of the ways to solve this problem is to employ a lattice with
extremely low momentum compaction factor (LMC) (see \cite{posipol05}
and the bibliography therein). Another known way  is to make use of
the longitudinal strong focusing lattice (LSF) with a sufficiently
high momentum compaction factor and a high RF voltage (see
\cite{gallo03} and the bibliography therein). The idea of LSF
consists in modulation of the bunch length over ring's circumference
such that the minimal length realized at the interaction point (IP)
(naturally, the maximum is at the RF cavity).

\paragraph{To maximize yield} one has to compress the bunch both transversally and (at non head-on collisions) longitudinally.

\paragraph{Energy spread} in Compton sources is the main draw-back:
\begin{itemize}
\item Longitudinal one may be cured with a high rf voltage and/or a low momentum compaction lattice;
\item Transversal chromaticity together with the spread leads to unstable motion of circulating particles
\end{itemize}

\paragraph{Bunch length} is also important for the Compton rings: it should be as short as possible at the collision point (cp) to provide the maximal yield and as long as possible beyond cp to mitigate deteriorating interactions with the environment (walls and joints of the vacuum chamber, etc.).

One of the way to reach proper modulation of the bunch length is employment of the strong longitudinal focusing scheme \cite{gallo03} (also see \cite{gallo03a}, \cite{gallo04}).

\begin{figure} [hbtp]
\centering
\includegraphics[width=\columnwidth]{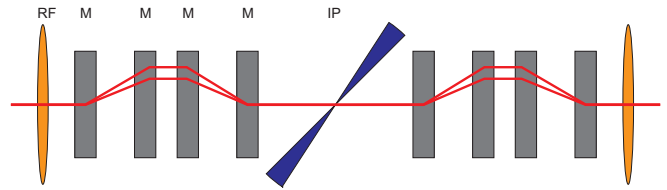}%
\caption[]{Scheme of the longitudinal low--$\beta $ insertion. RF
denotes radio--frequency cavities, M the rectangular bending
magnets, IP the interaction point(s) \label{chicane0}}
\end{figure}

\begin{figure} [hbtp]
\centering
\includegraphics[width=0.8\columnwidth]{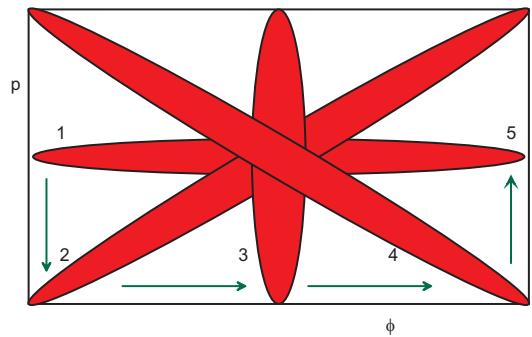}%
\caption[]{Transformation of bunch's longitudinal phase portrait. 1,
5 outside of LLBI, 2 at the first RF output, 3 at IP, 4 at the
second RF input \label{chicane1}}
\end{figure}

The essence of the idea of double chicane scheme is following: In
the low momentum compaction scheme the `transparent' insertion with
high momentum compaction (and high RF voltage as well) is
introduced. We will refer to this scheme as \emph{the longitudinal
low--$\beta$ insertion} (LLBI) in analog to the colliders
(transversal) low--$\beta$ insertions. A scheme of LLBI is presented
in Fig.~\ref{chicane0}, a (na\"ive) principle of operation depicted
in Fig.~\ref{chicane1}.

The principle of LLBI operation is as follows. A long bunch with a
small energy spread enters the RF cavity where the particles in the
head of the bunch have been decelerated while those in the tail
accelerated. Since the pass length in the chicane formed from the
rectangular magnets decreases with energy, the bunch reaches IP with
the minimal length. The second chicane and RF cavity, symmetric to
the first, restores the initial particle distribution.

In principle, the double chicane scheme represents the high--disperse ring with a single rf cavity where the bunch length undergoes modulation such that minimal length  attained at the mid--drift, the maximal at the mid--cavity, \cite{gallo03}. Formally LLBI reminds this scheme with the rf cavity divided into the two and the entire ring except for chicanes and cp placed within the rf halves.  Orientation of the bunch phase portrait 1,5 in Fig.\ref{chicane1} takes place over the rgular part of ring's orbit (in \cite{gallo03} -- only in the mid-cavity position).    

\section{Model of longitudinal beam dynamics in Compton rings}
In the model, a particle is being tracked on the phase space with the coordinate axes: phase $\phi $ relative to zero voltage in the rf cavity, and deviation of the reduced energy from the synchronous, $p=\delta\gamma /
\gamma_s $. Reduced time $\tau $ is counted in units of turns of the particle over the ring: $\tau = t \beta c /L$ ($L$ is the length of the orbit,
$\beta\equiv v/c$ the reduced velocity of the synchronous particle).

Transformations of the coordinates in the lattice elements reads:
\begin{itemize}
\item drift
\begin{eqnarray} \label{eq:drift}
\phi_f &=& \phi_i +\kappa \tau_\mathrm{dr} p_i\; , \nonumber \\
p_f &=& p_i\; ;
\end{eqnarray}

\item rf cavity
\begin{eqnarray}\label{eq:rf}
\phi_f &=& \phi_i \; , \nonumber \\
p_f &=& p_i-U_{RF}\sin\phi_i \; ;
\end{eqnarray}

\item collision point
\begin{eqnarray}\label{eq:ip}
\phi_f &=& \phi_i \; , \nonumber \\
p_f &=& p_i\left[ 1-b\zeta (2+ p_i)\right] -  b\zeta \; .
\end{eqnarray}

\end{itemize}

Here the following definitions accepted:
\begin{eqnarray*}
\kappa &=& 2\pi h \eta\; ; \\
b&=&2(1+\cos\varphi ) \gamma_s E_\mathrm{las}/E_0\; ; \\
U_{RF} &=& eV/(\gamma_s E_0)\; ,
\end{eqnarray*}
with $h$ the harmonic number (ratio of the orbit length to rf wavelength); $\tau_\mathrm{dr} $ ratio of the drift length to the orbit length; $V$ rf voltage amplitude; $\eta $ the momentum compaction factor; $\zeta $ ratio of the energy of scattered (Compton) photon to its maximal value.
\section{Linear model. Analysis.}
Let us restrict our consideration to a linear case, $\sin \phi \approx \phi$, without Compton interactions, $b = 0$.

\subsection{Small azimuthal variations.}
With small azimuthal variations over turn, $|\phi_f - \phi_i|=\delta\ll 1$, $|p_f - p_i|=\epsilon\ll 1$: low rf voltage and small momentum compaction factor,  Eqs. \eqref{eq:drift} and \eqref{eq:rf} are connected to Hamilton function
\begin{equation} \label{eq:hamilt}
H = \kappa p^2/2 + U_{RF}\cos \phi\;,
\end{equation}
which represents mathematical pendulum with frequency of small oscillations $\Omega^2\approx U_{RF}/\kappa$.

\subsection{Large azimuthal variations. Single cavity.}
With a high--dispersion drift and single rf cavity, strong longitudinal focusing pattern is realized, see \cite{gallo03a,gallo04,biscari05,falbo06}.

In this case, longitudinal motion can be treated similar to transverse motion in strong--focusing lattice of FOF/OFO  type. Transformation of $(\phi,p)$ vector described by \eqref{eq:drift}, \eqref{eq:rf} can be represented by matrices
\begin{gather}\label{eq:matrices}
\mathbf{M}_\mathrm{drift} =
\begin{pmatrix} 1 & \kappa\\
0& 1
\end{pmatrix}\; , \qquad
\mathbf{M}_\mathrm{rf} =
\begin{pmatrix} 1 & 0\\
-U& 1
\end{pmatrix}\; .
\end{gather}

Let us write the full--turn cycle matrix as consisted of two half--drifts $(\tau = 1/2+1/2)$ and two halves of rf cavity $(U/2+U/2)$
\begin{align} \label{eq:matcycle}
\mathbf{M}_\mathrm{min} &= \mathbf{M}_\mathrm{drift/2} \cdot \mathbf{M}_{U/2} \cdot \mathbf{M}_{U/2} \cdot \mathbf{M}_\mathrm{drift/2}\; ;\\
\mathbf{M}_\mathrm{max} &= \mathbf{M}_{U/2} \cdot \mathbf{M}_\mathrm{drift/2}  \cdot \mathbf{M}_\mathrm{drift/2} \cdot \mathbf{M}_{U/2}\; .
\end{align}

From any of these matrices, one can get the stability boundaries
\begin{equation} \label{eq:stab}
\left| \mathrm{Tr}\mathbf{M}\right| = \left| 2 - \kappa U \right|\le 2\; ,
\end{equation}
Thus, the stable motion can exist if
\begin{equation} \label{eq:stab1}
0 \le \kappa U \le 4\; .
\end{equation}

The left limit corresponds to very weak focusing (zero frequency of the synchrotron oscillations), the right one to very strong focusing (complete synchrotron oscillation lasts two turns).

In general, in non-uniform azimuthal focusing amplitude of synchrotron oscillations (and the bunch length) vary along the orbit. Modulation of the amplitude -- squared ratio of minimal amplitude (in the center of the drift) to maximal one (in the median section of rf cavity) -- equal to ratio of the matrix elements
\begin{equation} \label{eq:modul}
\mathrm{modulation} = \frac{M^\mathrm{(min)}_{12}}{M^\mathrm{(max)}_{12}} = 1 - \frac{\kappa U}{4}\; .
\end{equation}

As expected, minimal modulation takes place for the weak focusing ($ 0 \leftarrow\kappa U$), it is rose to full modulation at the right boundary of stability  ($\kappa U\to 4$).

\subsection{Large azimuthal variations. Two cavities.}
Making use similar procedure, we treat a lattice comprising two rf cavities separated with two drifts. The drifts may possess different chromaticity (phase slip factors $\kappa_{1,2}$). This scheme represents OFOFO or OFODO pattern of focusing depending on rf cavities in-phase or in inverted phases.

Most promising case is of identical phased cavities, $U=V$, (OFOFO type). Trace of the cyclic matrix is
\begin{equation} \label{eq:trfof}
\mathrm{Tr} \mathbf{M}_\mathrm{fof} = 2 - 2(\kappa_{1}+\kappa_{2})U + \kappa_{1}\kappa_{2}U^2\; ,
\end{equation}
and stability region is symmetric with respect to $\kappa_{1}+\kappa_{2}$. Designate $k=\kappa_1 U$ and $q=\kappa_2 U$, we get the stability region limited with
\begin{align} \label{eq:trfof1}
-\frac{2q}{2-q}\le &k \le 2\; ; & -\frac{2k}{2-k}\le &q \le 2\;  .
\end{align}

\begin{figure} 
\centering
\includegraphics[width=0.8\columnwidth,height=0.75\columnwidth]{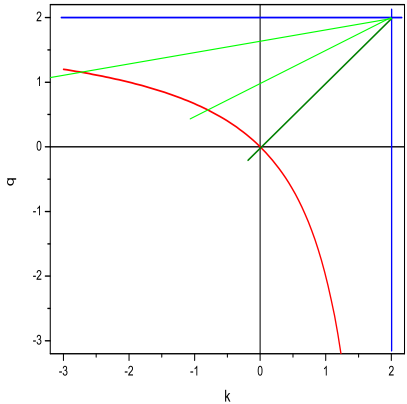}%
\caption[]{Stability region (red and blue curves) and lines of equal modulation (green). \label{fig:stabregion}}
\end{figure}

The stability region is limited from top-right with straight lines (right angle) $q=2$, $k=2$, and hyperbola $kq-2(k+q)=0$ from left-bottom. With one of the parameters $(k,q)\to -\infty$, the interval of stability for another becomes narrower as $2-4/q\le k\le 2, p\to -\infty$ (see Fig.\ref{fig:stabregion}).

The $M_{12}$ element of the cyclic matrix at symmetric point -- center of the $\kappa_1$ drift (equal to $\beta\sin\mu$) -- reads
\begin{equation} \label{eq:m12fof}
M_{12}(\tau = \kappa_1/2) = \frac{1}{4}(2-\kappa_1U)[2(\kappa_1+\kappa_2)-\kappa_1\kappa_2U]\;.
\end{equation}

From \eqref{eq:m12fof} it follows the ratio of the magnitudes of enveloping function in mid-drifts:
\begin{equation} \label{eq:modul1}
\xi \equiv\frac{M_{12}(\tau = \kappa_1/2)}{M_{12}(\tau = \kappa_2/2)} = \frac{2-\kappa_1U}{2-\kappa_2U} = \frac{2-k}{2-q}\; .
\end{equation}

Combining \eqref{eq:trfof} and \eqref{eq:modul}, we can derive the stability extend for the given modulation $\xi$:
\begin{align} \label{eq:stabxi}
&0\le 2-q\le 2\left/\sqrt{\xi}\right. \; , &  &0\le 2-k\le 2\sqrt{\xi}\; .
\end{align}

\paragraph{Stability region limited by sinusoidal rf}
From \eqref{eq:modul1} with account \eqref{eq:trfof1} it may follow that to get maximal modulation of the bunch length one must put $k\lesssim 2$ and $q\to -\infty$ if the bunch survives in a narrow stable region. But actually, sinusoidal rf voltage \eqref{eq:rf} puts the limits on the parameters. Really, maximal kick induced by the cavity at $\phi = \pm \pi / 2$ should cause phase change $\Delta\phi \le \pi$. Therefore
\begin{align} \label{eq:nonlstab}
&p_f = - p_i\;; & &\Delta \phi_\mathrm{max} = \pi\; ;\nonumber \\
&|p|_\mathrm{max} = U/2\; ; & &|\kappa U|= |k| \le  2 \pi \; .
\end{align}

There is a more serious consequence from this estimation: comparing the stable interval of strong focusing \eqref{eq:nonlstab} with that of weak focusing from \eqref{eq:hamilt}, one can see the weak focusing island of stability, $-\pi\le \phi \le \pi$, is two times wider than the strong focusing one \eqref{eq:nonlstab}, $-\pi/2\le \phi \le \pi/2$.

\section{Simulations}
\begin{figure} 
\centering
\includegraphics[width=0.8\columnwidth]{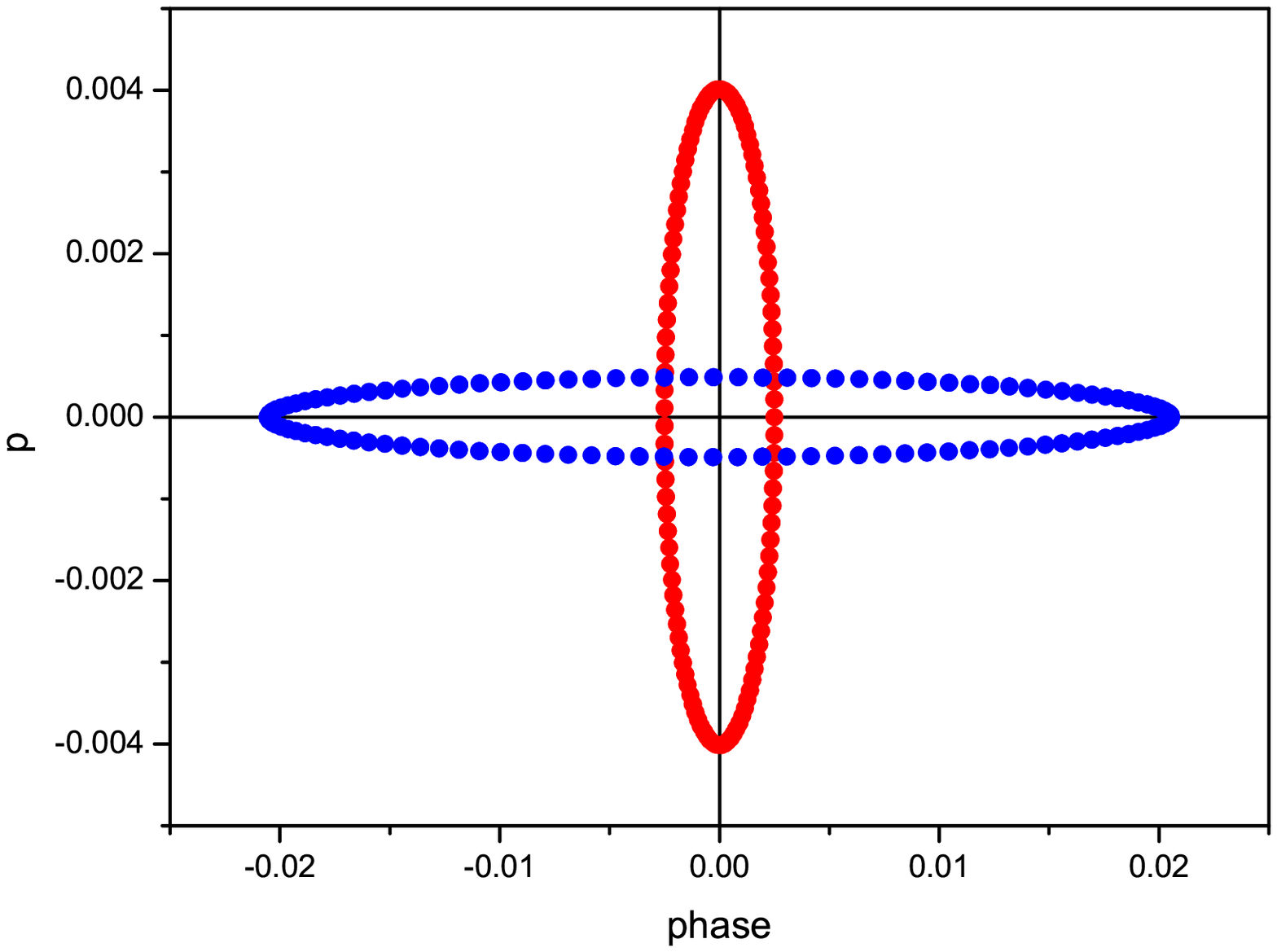}\\
\includegraphics[width=0.8\columnwidth]{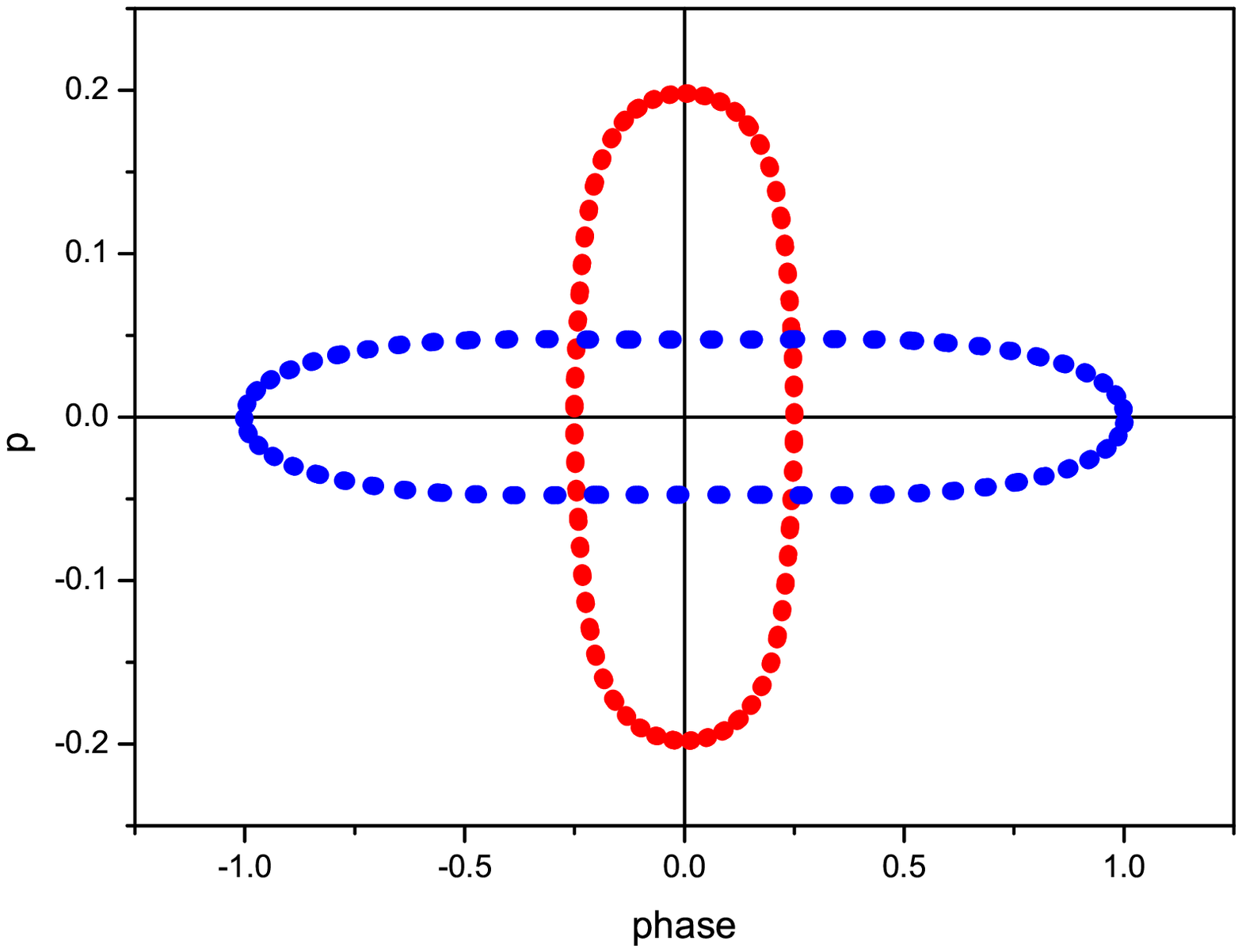}
\caption[]{Phase trajectories in the case $2- k = 0.062\;, q =-2.22$ (red color at CP, blue at mid drift). The top graph presents small initial deviation, the bottom one large deviations.  \label{modlin}}
\end{figure}

\paragraph{Phase trajectories} were simulated for a case of absence both synchrotron damping and Compton interactions. With two-cavity lattice, phase trajectories were registered at the collision point (in between chicanes and at the opposite point (middle of the drift).
In Fig.\ref{modlin}, there presented are the phase trajectories at CP and at mid drift, for the initial deviations of simulated particles resembling the linear motion (small initial deviation) and the nonlinear one (large deviation).

As it seen from the trajectories, amplitude of the envelope of phase oscillation at CP is much smaller then at mid-drift, for the envelope of energy deviation opposite situation holds.
It worth to mention that modulation of the phase oscillations envelope in the simulation shows fairly good agreement with the theoretical prediction,
\[
\frac{\phi^\mathrm{(max)}_\mathrm{cp}}{\phi^\mathrm{(max)}_\mathrm{md}} = \sqrt{\frac{2-k}{2-q}}\approx 0.122\; .
\]
With increasing impact of nonlinearity, the modulation decreased (see Fig.\ref{modlin} the bottom panel, modulation $approx 0.25$), trajectories become nonelliptic.

\begin{figure} 
\centering
\includegraphics[width=0.8\columnwidth]{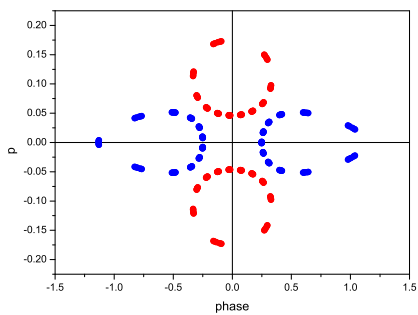}%
\caption[]{Phase trajectories in the case $\mathrm{sign}\; k =  \mathrm{sign}\; q$ (blue color at CP, red at mid drift). \label{fofo3}}
\end{figure}

The nonlinear trajectories are presented in Fig.\ref{fofo3} for the case of the same signs of the phase slip value.

Each trajectory represents two--loop trace (for the single particle!), orientation inverse to the case of opposite signs.

\subsection{Simulation of CLIC-like ring}
Simulation of a Compton ring dedicated for generation of polarized gammas in continual regime show advantages of the two--chicane scheme with opposite signs of  the phase slip factors and $q \approx -k$. Simulation revealed stability of the scheme (no losses) and high yield of gammas as compared with the scheme with  $\mathrm{sign}\; q = \mathrm{sign}\; k$, $q\ll k$, see [PosiPol~09].

\begin{figure} 
\centering
\includegraphics[width=0.8\columnwidth]{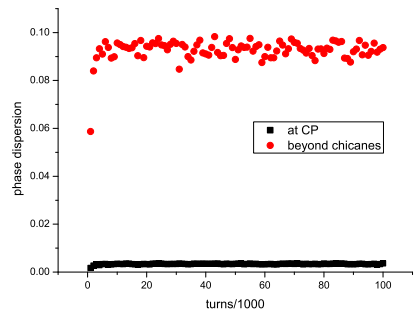}%
\caption[]{Mean squared bunch phase length vs. time. \label{chiclength}}
\end{figure}

\begin{figure} 
\centering
\includegraphics[width=0.8\columnwidth]{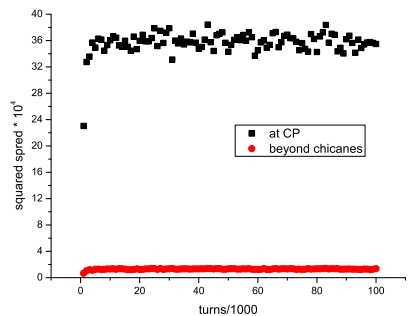}%
\caption[]{Mean squared energy spread vs. time. \label{chicspread}}
\end{figure}

In Fig.\ref{chiclength} and Fig.\ref{chicspread} the mean squared bunch length and momentum spread are plotted.

\subsection{For further consideration: Compensation of $\sin \phi$ nonlinearity by $\kappa_3 p^3$ momentum compaction nonlinearity}

\section{Summary and Outlook}
The single-particle dynamics of electron bunches in Compton gamma
sources has been studied analytically and in simulations. The
analytical estimates for the steady-state bunch parameters are in
good agreement with those obtained from the simulations.

Our study suggests that the interaction of the electrons with the
laser photons do not significantly affect the transverse degrees of
freedom. The `bottleneck' is the longitudinal dynamics: the ring's
energy acceptance should be unusually large.

%

\end{document}